\documentclass[twocolumn,superscriptaddress]{revtex4-1}
\usepackage[T1]{fontenc}
\usepackage[utf8]{inputenc}
\usepackage[english]{babel}
\usepackage{graphicx}
\usepackage{amsmath,amssymb}
\usepackage{mathrsfs}
\usepackage{booktabs}
\usepackage{bm}
\usepackage{hyperref}

\begin{document}

\title{Quantum Monte Carlo study of a two dimensional dipolar Bose gas in a harmonic trap}
\date{\today}
\author{Michele Ruggeri}
\email{michele.ruggeri@sissa.it}
\affiliation{SISSA Scuola Internazionale Superiore di Studi Avanzati and DEMOCRITOS National Simulation Center, Istituto Officina dei Materiali del CNR Via Bonomea 265, I-34136, Trieste, Italy}

\begin{abstract}

We present a Quantum Monte Carlo study of the ground state properties of a two dimensional system of Bose particle with dipole moment in a harmonic trap. The direction of the dipoles is assumed to be fixed by an external field. We study how the system behaves when the direction of the dipole moments is changed. Our analysis is made using Path Integral Monte Carlo simulations in the Grand Canonical ensemble, using the Worm Algorithm. We study systems of increasing size, focusing on the spatial distribution of the atoms inside the trap and on the superfluid fraction of the sample. We find that vertical dipoles form a crystal, characterized by a shell structure, while tilting the dipoles has the effect of inducing a striped structure. We find that similar behaviour can be found in the analogue classical system. Vertical dipoles display nonclassical inertia while a strong tilting decreases superfluidity.
\end{abstract}

\maketitle
\section{Introduction}
Recently there have been several studies in the field of ultra cold quantum gases with large magnetic dipoles\cite{pollet_review} \cite{stringari_review}; for example Dy atoms were cooled and trapped \cite{dis}, and Bose-Einstein condensation has been observed in a $^{52}$Cr system \cite{bec_cr_1}\cite{bec_cr_2}. Also studies on molecules with electric dipole such as $^{40}$K$^{87}$Cr have been made (see e.g.\ \cite{molec}).
The interaction between dipoles is especially interesting for two reasons: it has long range and it is anisotropic. In ultra cold quantum gases in fact there are usually only s-wave contact interactions, which are short ranged and isotropic, and adding dipolar interaction bring novel features to such systems.
 
The anisotropy of dipolar interactions has two main consequences: the appearance of instability in 3D systems and the formation of several peculiar structures.
Large dipolar systems are unstable in three dimensions, due to the head to tail attraction \cite{biconc,instability}; in order to avoid this problem pseudo-2D systems, obtained with anisotropic optical traps, are studied: in a reduced dimensionality the system becomes stable, provided the interaction is sufficiently repulsive \cite{2dstab}. 
The anisotropy has also been shown to be the cause of remarkable structures in ultra cold gases. An example is the biconcave condensate described in \cite{biconc}, or the stripes that appear when the dipoles are tilted \cite{boronat_bulk,giorgini_fermi}.

In this work we study a two dimensional system of Bose atoms with dipole moment in a harmonic trap. The dipole moments are all parallel, and their direction is assumed to be fixed by an external magnetic or electric field. Our aim in particular is to study the influence of inclination of the dipoles with respect to the $xy$ plane on the physical properties of the system. We note in fact that while 2D or pseudo-2D systems have also been studied before (see for example \cite{instability, 2dstab, zoller_rev, reimann_diag, reimann_curr, boronat_bulk, boronat_bulk_2, giorgini_fermi, boninsegni_dipoles, boninsegni_2, roton, gads_bruun1, gads_bruun2}) most of this studies were made either for bulk systems or for trapped systems with vertical dipoles.

This physical system was previously studied via exact diagonalization in \cite{reimann_diag}; of course exact diagonalization only allows the study of very small systems: only a 3 particles systems could be studied. We instead use a Quantum Monte Carlo technique, that allows us to consider larger systems; in particular we will study a system made by 3, 19 and 100 atoms.

We focus on the structural properties of the dipolar system, studying the single particle density and the radial pair correlation function, to see how the tilt angle affects them. We compare the structures emerging in the quantum system with the ones of the classical analogue, estimated with classical Monte Carlo simulations. We also evaluate the superfluid fraction and the total energy of our sample.

This document is organized as follows.
In Section II we discuss the physical system in analysis. In Section III we describe the simulation technique we used. In Section IV we show the results of our simulations and finally we present our conclusions in Section V.

\section{Physical system}
We consider a system of $N$ atoms in the $xy$ plane in a optical trap; these atoms interact because of thei dipole moment.
The Hamiltonian of our system is
\begin{equation}
H = \sum_i \frac{p_i^2}{2m_i} + \sum_i V_{trap}({\bf r}_i) + \frac{1}{2}\sum_{i\neq j}V_{int}({\bf r}_i-{\bf r}_j).
\end{equation}
$V_{trap}$ here represents an optical trap like the ones used in experiments on ultra cold quantum gases. Such a trap in general can be modelled using a harmonic potential of the form
\begin{equation}
V_{trap}({\bf r}) = \frac{1}{2}m\left( \omega_x^2 x^2+\omega_y^2 y^2+\omega_z^2 z^2\right).
\label{trap}
\end{equation}
The interaction between the atoms in the trap is due to their dipole moments. The dipole moments are vectors  in the $xz$ plane; they are all parallel, and their direction is given by a strong external magnetic or electric field. Following the notation in \cite{reimann_diag} we define the tilt angle $\Theta$ as the angle between the moments and the $x$ axis; for reference $\Theta = 90^\circ$ means that the dipoles are aligned to the $z$ axis, while tilting the dipoles means decreasing $\Theta$. 
The interaction between two parallel dipoles can be written as
\begin{equation}
V_{int}({\bf r})=D^2\frac{1-3\cos^2\theta_{rd}}{r^3};
\label{pot}
\end{equation}
here $D$ is the dipole strength and $\theta_{rd}$ is the angle between the dipole moment and the vector $\bf r$. If our atoms have an electric dipole $d$ we have $D^2=\frac{d^2}{4\pi\epsilon_0}$, while if they have a magnetic dipole $d$ we have $D^2=\mu_0 d^2/\left(4\pi\right)$. 

It has been shown that a system composed by dipoles is unstable in three dimensions \cite{instability}, so experiments are usually made in a pseudo 2D environment, obtained using a strongly anisotropic trap. This means that in the equation (\ref{trap}) we have $\omega_z \gg \omega_x = \omega_y = \omega_0$; defining the oscillator length
\begin{equation}
l=\sqrt{\frac{\hbar}{m\omega}}
\end{equation}
this means that $l_z \ll l_0$. In the computations we can make a choice: we can consider a pseudo 2D system, using an effective potential, obtained integrating the actual potential along the $z$ direction as in \cite{reimann_diag,reimann_curr} or we can consider a strictly 2D system, i.e. taking the limit $\omega_z \rightarrow \infty$ ($l_z \rightarrow 0$). We saw after performing some preliminary computations that if we set a $l_z$ small enough the results for the two systems are essentially the same; in order to speed up the computations we decided to focus on the strictly 2D system, for which we can use the potential \ref{pot}.
In 2D the potential \ref{pot} is repulsive and isotropic as long as $\Theta=90^\circ$; tilting the dipoles has the effect of causing anisotropy, and along with anisotropy head to tail attraction appears.  
Due to this attraction, decreasing $\Theta$ too much will cause the system to collapse. The 2D system however has been shown to be stable for $\Theta > \arccos \frac{1}{\sqrt{3}} \simeq 54.7^\circ$ \cite{instability,reimann_diag,reimann_curr}. In order to avoid to deal with the instability we performed our computations in the region $55^\circ \leqslant \Theta \leqslant 90^\circ$.

\section{Simulation method}

Quantum Monte Carlo techniques are widely used in the study of condensed matter systems, and in particular they have several applications in the field of ultra cold quantum gases \cite{pollet_review}. An especially useful technique for such systems is the Worm algorithm \cite{worm}. This algorithm, which is a variant of the Path Integral Monte Carlo scheme, allows the exact evaluation of thermal averages of physical observables for a system of Bose particles in the Grand Canonical ensemble at finite temperature. 

In Path Integral Monte Carlo simulations \cite{ceperley_4he} the quantum classical isomorphism is used, so that it is possible to study a quantum system by simulating a classical system of special polymers. The Worm algorithm allows to easily take into account the interparticle permutations that have to be introduced when studying a system of indistinguishable quantum particles. Further details on these techniques can be found in \cite{ceperley_4he,worm}. We note that Path Integral Monte Carlo suffers from the sign problem when dealing with fermions: for this reason we focus only on bosons.

In all our simulations we set $\hbar=\omega_0=m=k_B=1$, so that we have $l_0=1$. Having made this choice we measure all the lengths in units $l_0$ and all the energies and temperatures in units $\hbar\omega_0$. We set the imaginary time increment $\tau=2\cdot 10^{-2}$. For the propagator we used the primitive approximation. All our simulations were made at the temperature $T=0.1$. We made this choice because we were interested in studying the ground state behaviour of our system; previous works \cite{reimann_diag} shown that the energy gap between the ground state and the first excited state should be at least $0.5$ in our energy units, so using $T=0.1$ should allow us to avoid meaningful contributions of the excited states to our thermal averages. Using a finite temperature technique to study ground state properties may seem odd, but it should be noted that our technique unlike zero temperature QMC schemes does not use trial wave functions that may introduce a bias in the computations. In accord to \cite{reimann_diag} we set $D=5.0$ in all the simulations.  We considered systems of increasing size, starting from $N=3$ particles and then going to $N=19$ and $N=100$. Using the Worm algorithm we are working in the Grand Canonical ensemble, and so the particle number fluctuates; its average value is set by the chemical potential $\mu$ of the simulation. In our simulations the fluctuations of this quantity are practically negligible, so in practice we can think we are working with a fixed particle number.

The physical quantities we evaluate are the single particle density $\rho(x,y)$, the radial correlation function $g(r)$, the energy per particle $E$ and the superfluid fraction $\rho_S/\rho$. We perform several computations, varying the tilt angle $\Theta$ from $55^\circ$ (near to the stability threshold) to $90^\circ$ (vertical dipoles). 

The density profile $\rho(x,y)$ is obtained simply by evaluating the histogram of the position of the particles during the simulation. Similarly the radial correlation function $g(r)$ is estimated by taking the histogram of the distances of the particle pairs. Since we are considering a finite system we are not using periodic boundary conditions, and so we can't use the winding number estimator to compute the superfluid fraction; instead, we use the area estimator \cite{ceperley_4he,ceperley_cluster}.

Along with the Quantum Monte Carlo simulations we also performed a classical study of the trapped dipoles system. This classical study consisted firstly in determining the classical ground state of our system. In order to do that we used the simulated annealing technique \cite{annealing}. 
Then we made classical Monte Carlo simulations at finite temperature; we made these computations to compare $\rho(x,y)$ in the quantum and classical cases, so that we can compare the effect of thermal and quantum fluctuations on the properties of the trapped dipoles system.

\section{Results}
\begin{figure}
\includegraphics[angle=270,width=\columnwidth]{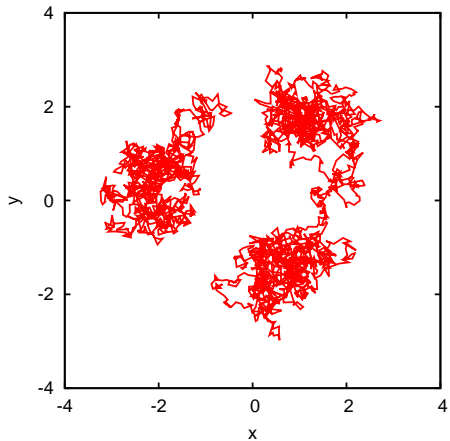}
\caption{\label{triang}Example of configuration of a three particle simulation with $\Theta=90^\circ$, $D=5.0$, $T=0.1$. Each point is the position of a particle in a given timeslice. A triangular structure can be seen. The lengths are in units of $l_0$.}
\end{figure}

\begin{figure}
\includegraphics[angle=270,width=0.32\columnwidth]{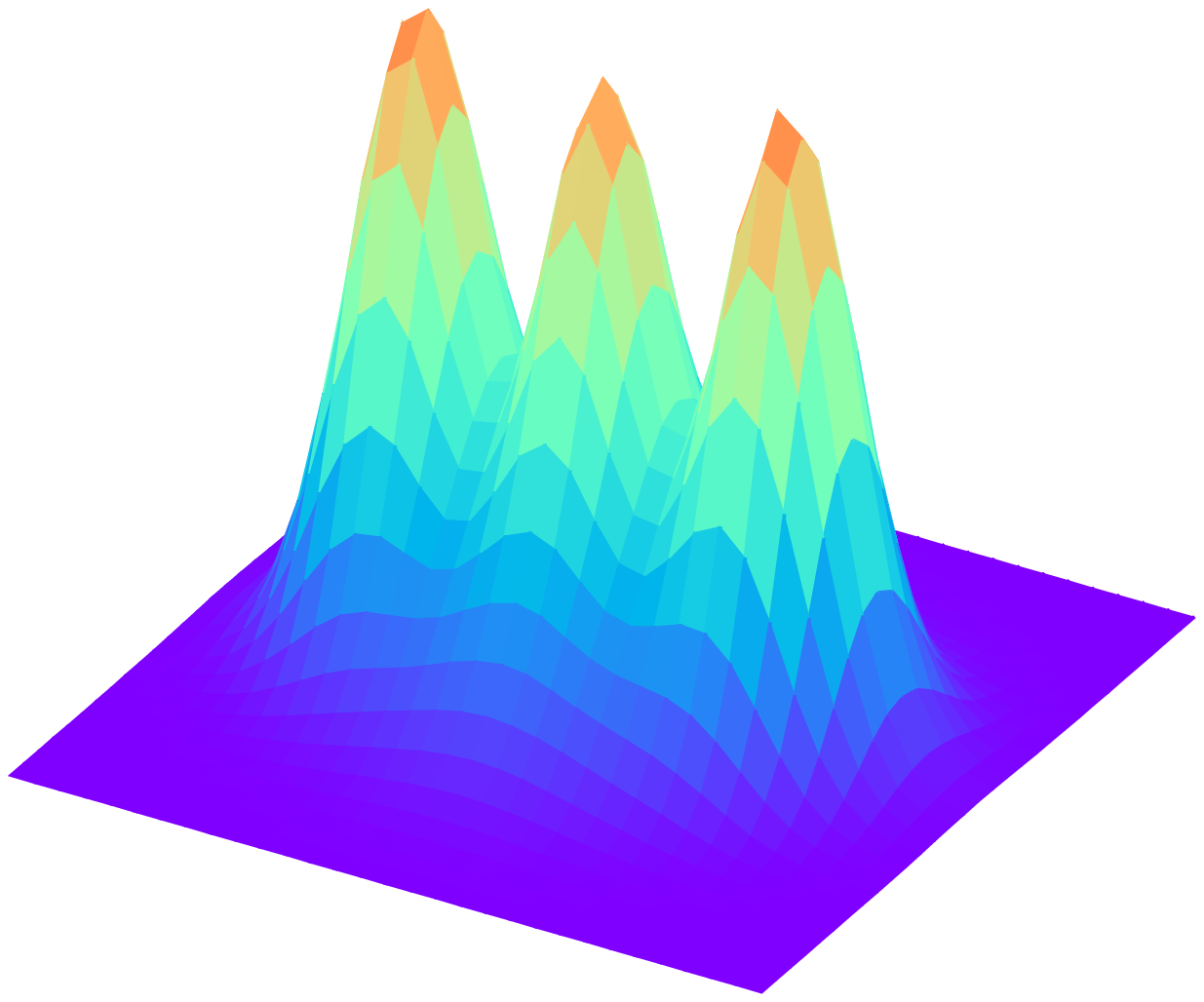}
\includegraphics[angle=270,width=0.32\columnwidth]{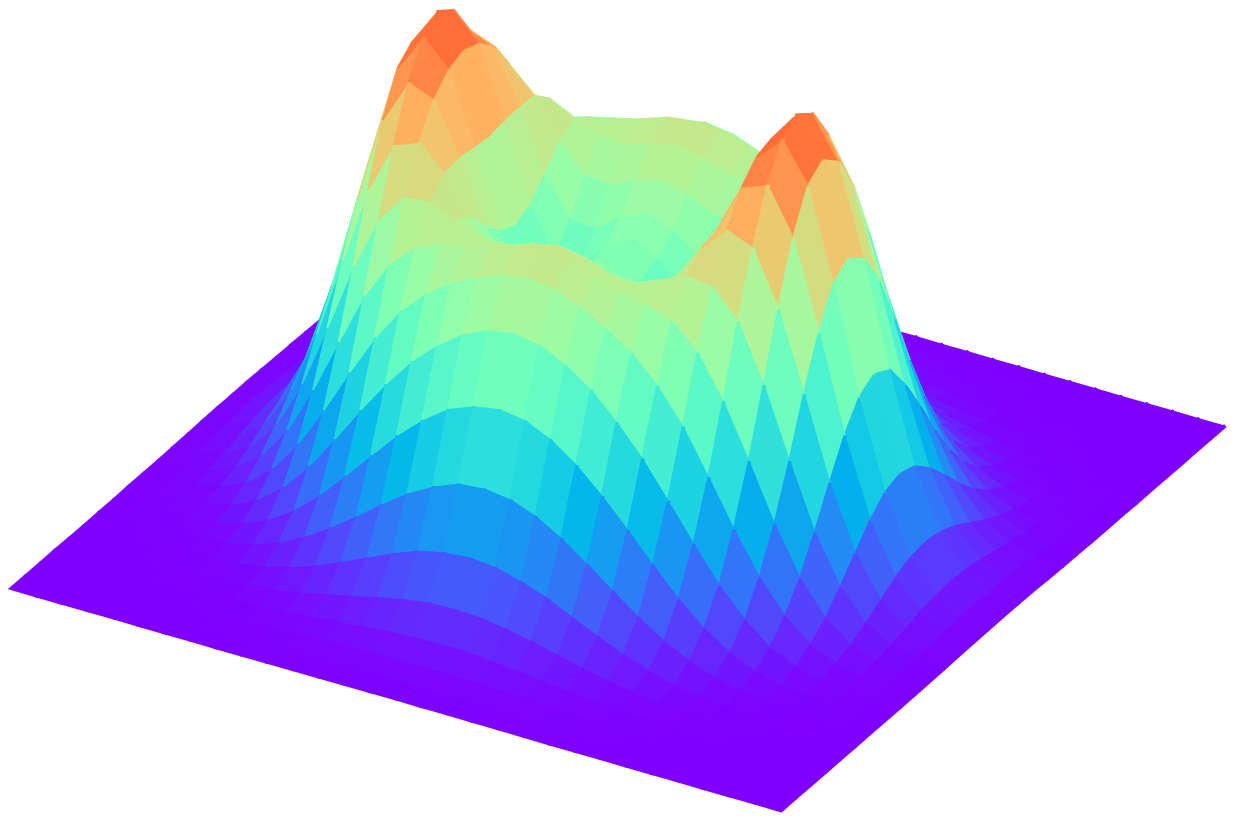}
\includegraphics[angle=270,width=0.32\columnwidth]{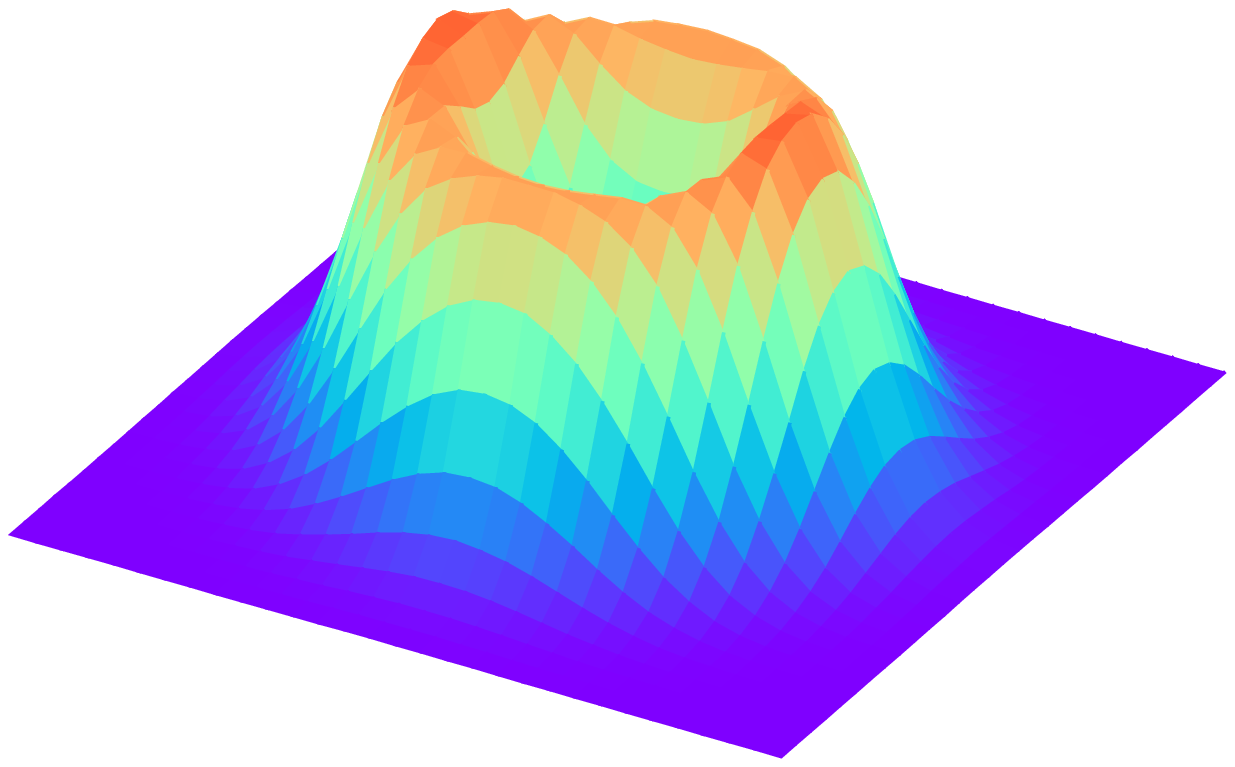}
\includegraphics[angle=270,width=0.32\columnwidth]{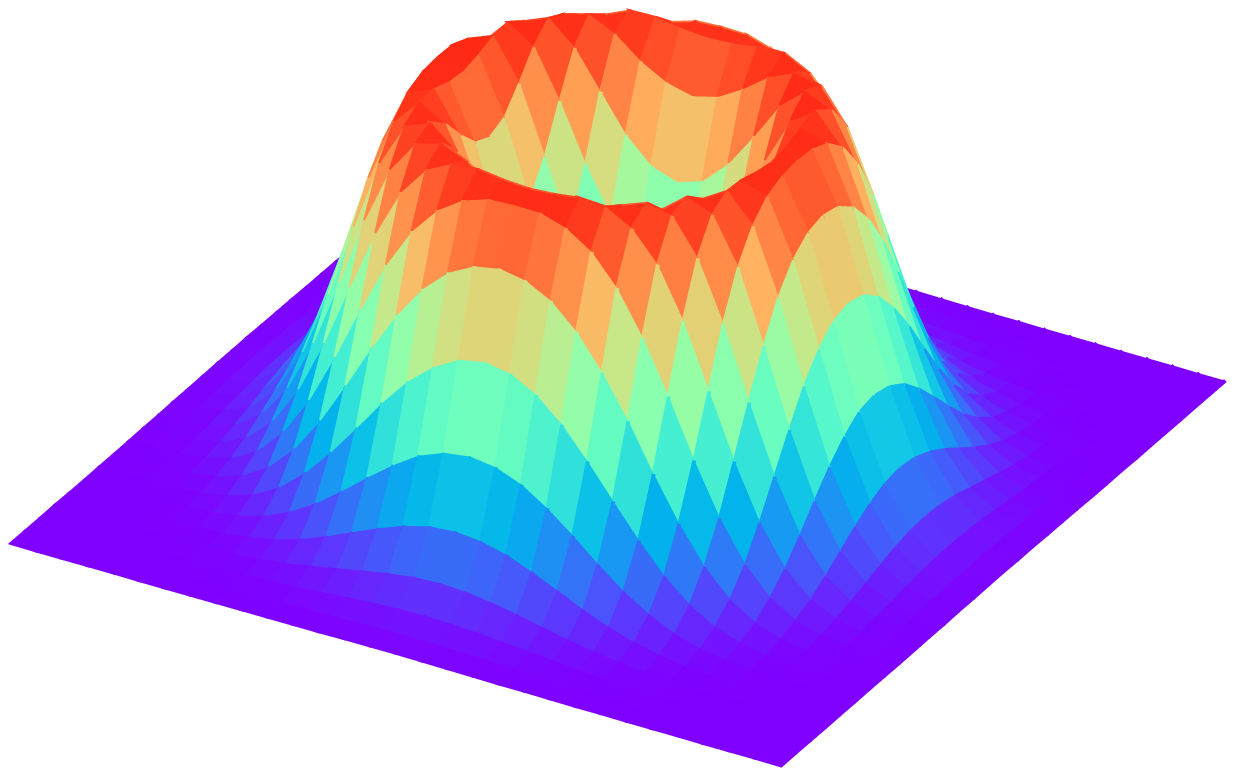}
\includegraphics[angle=270,width=0.32\columnwidth]{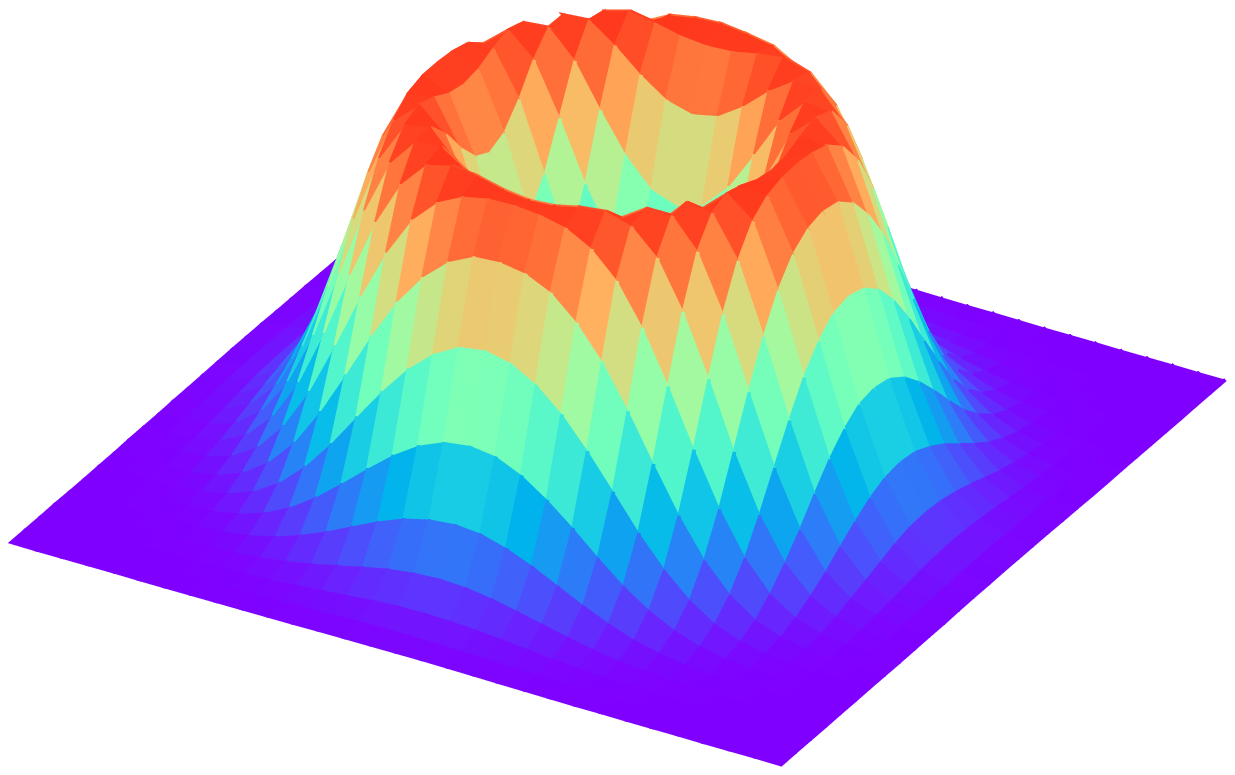}
\includegraphics[angle=270,width=0.32\columnwidth]{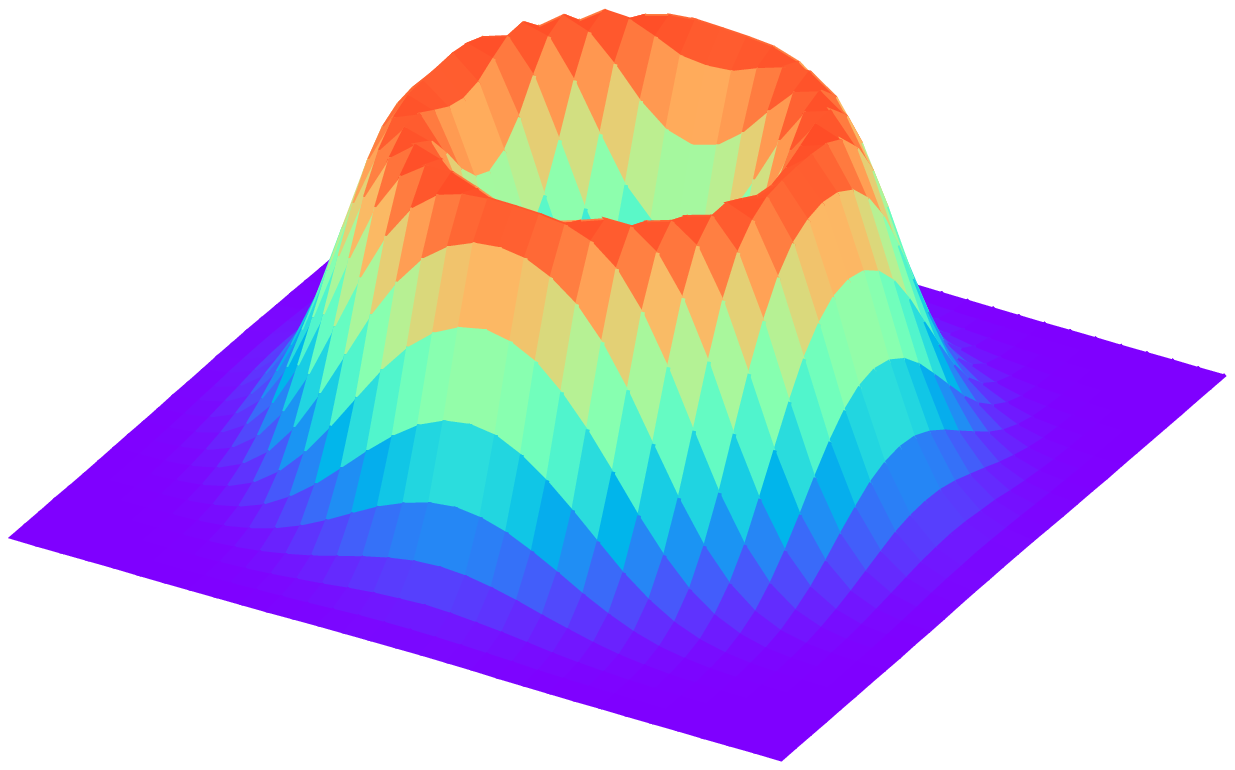}
\caption{\label{ro_3}Particle density for a three particle system with increasing $\Theta$. We have respectively $\Theta=55^\circ$, $\Theta=57^\circ$, $\Theta=59^\circ$, $\Theta=65^\circ$, $\Theta=70^\circ$, $\Theta=90^\circ$.}
\end{figure}

Firstly let us consider the structural properties of our system, i.e. the density and the radial correlation function. We start with a system of three dipoles. For vertical dipoles, $\Theta=90^\circ$, the interaction is isotropic and repulsive. In the ground state the dipoles form an equilateral triangle around the centre of the trap. We can see that by observing a snapshot of a simulation in figure (\ref{triang}).
This triangle during the simulation is free to rotate around the centre of the trap, so that on average the density displays a hole in the centre and a maximum at a fixed radius.
If we instead consider tilted dipoles this behaviour changes. A system with small $\Theta$, $\Theta\simeq 55 ^\circ$, displays a linear structure: the atoms form a line which is parallel to the $x$ axis. By increasing the tilt angle  $\Theta$ the linear structure is progressively lost, and the isotropic behaviour is fully restored for $\Theta\gtrsim60^\circ$. We show in figure (\ref{ro_3}) the density profiles for increasing tilt angles from $\Theta=55^\circ$ to $\Theta=90^\circ$. We note that our single particle densities are in good accord with the ground state results obtained with exact diagonalization \cite{reimann_diag}.

\begin{figure}
\includegraphics[angle=270,width=0.32\columnwidth]{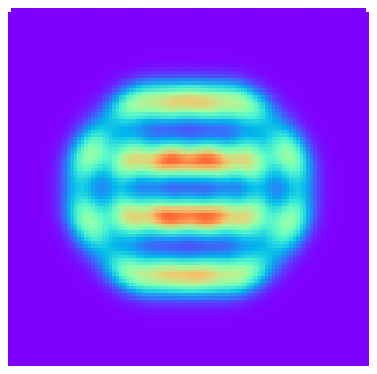}
\includegraphics[angle=270,width=0.32\columnwidth]{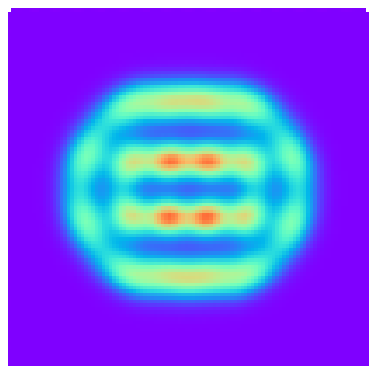}
\includegraphics[angle=270,width=0.32\columnwidth]{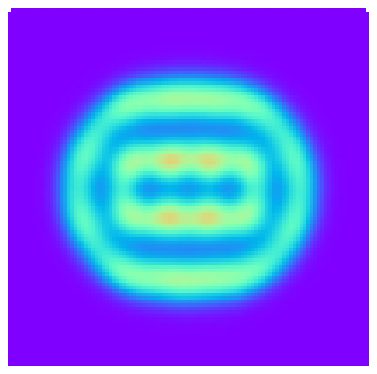}
\includegraphics[angle=270,width=0.32\columnwidth]{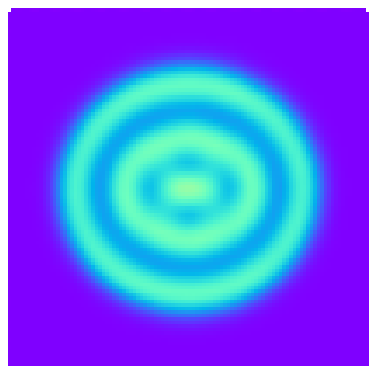}
\includegraphics[angle=270,width=0.32\columnwidth]{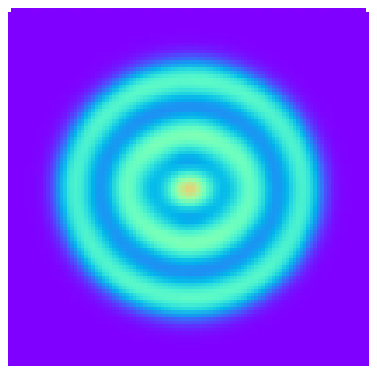}
\includegraphics[angle=270,width=0.32\columnwidth]{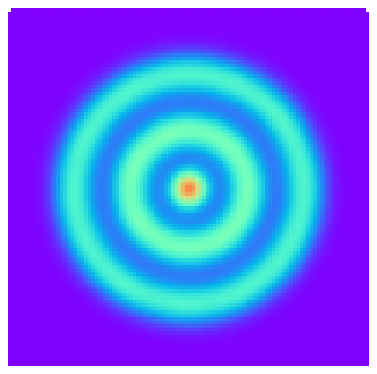}
\caption{\label{ro_19}Particle density for a 19 particle system with increasing $\Theta$. We have respectively $\Theta=55^\circ$, $\Theta=57^\circ$, $\Theta=59^\circ$, $\Theta=65^\circ$, $\Theta=70^\circ$, $\Theta=90^\circ$.}
\end{figure}

\begin{figure}
\includegraphics[angle=270,width=0.32\columnwidth]{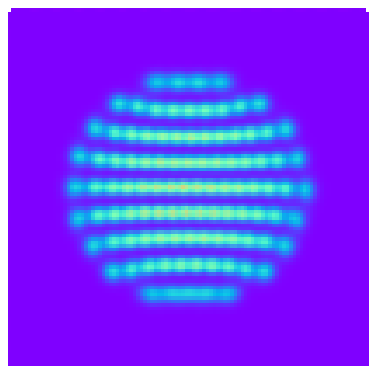}
\includegraphics[angle=270,width=0.32\columnwidth]{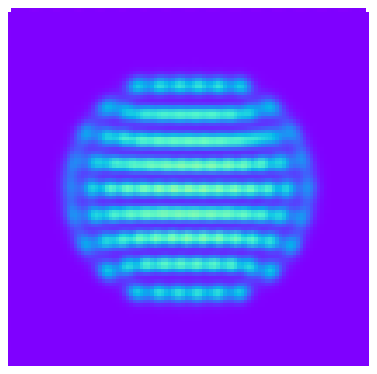}
\includegraphics[angle=270,width=0.32\columnwidth]{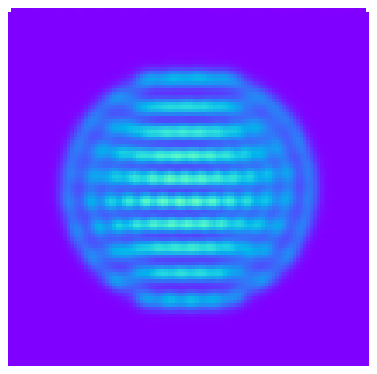}
\includegraphics[angle=270,width=0.32\columnwidth]{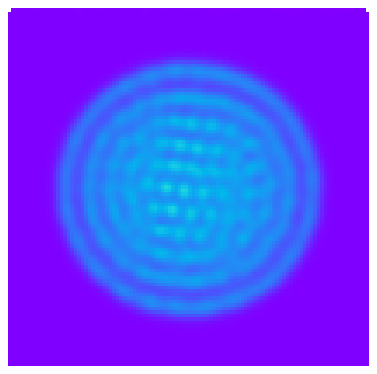}
\includegraphics[angle=270,width=0.32\columnwidth]{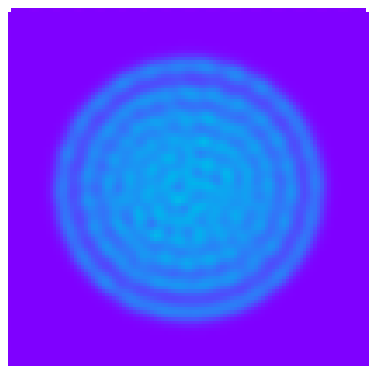}
\includegraphics[angle=270,width=0.32\columnwidth]{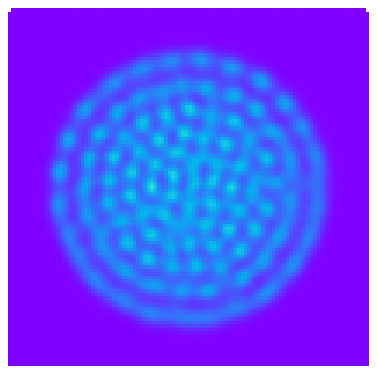}
\caption{\label{ro_100}Particle density for a 100 particle system with increasing $\Theta$. We have respectively $\Theta=55^\circ$, $\Theta=57^\circ$, $\Theta=59^\circ$, $\Theta=65^\circ$, $\Theta=70^\circ$, $\Theta=90^\circ$.}
\end{figure}

\begin{figure}
\includegraphics[angle=270,width=\columnwidth]{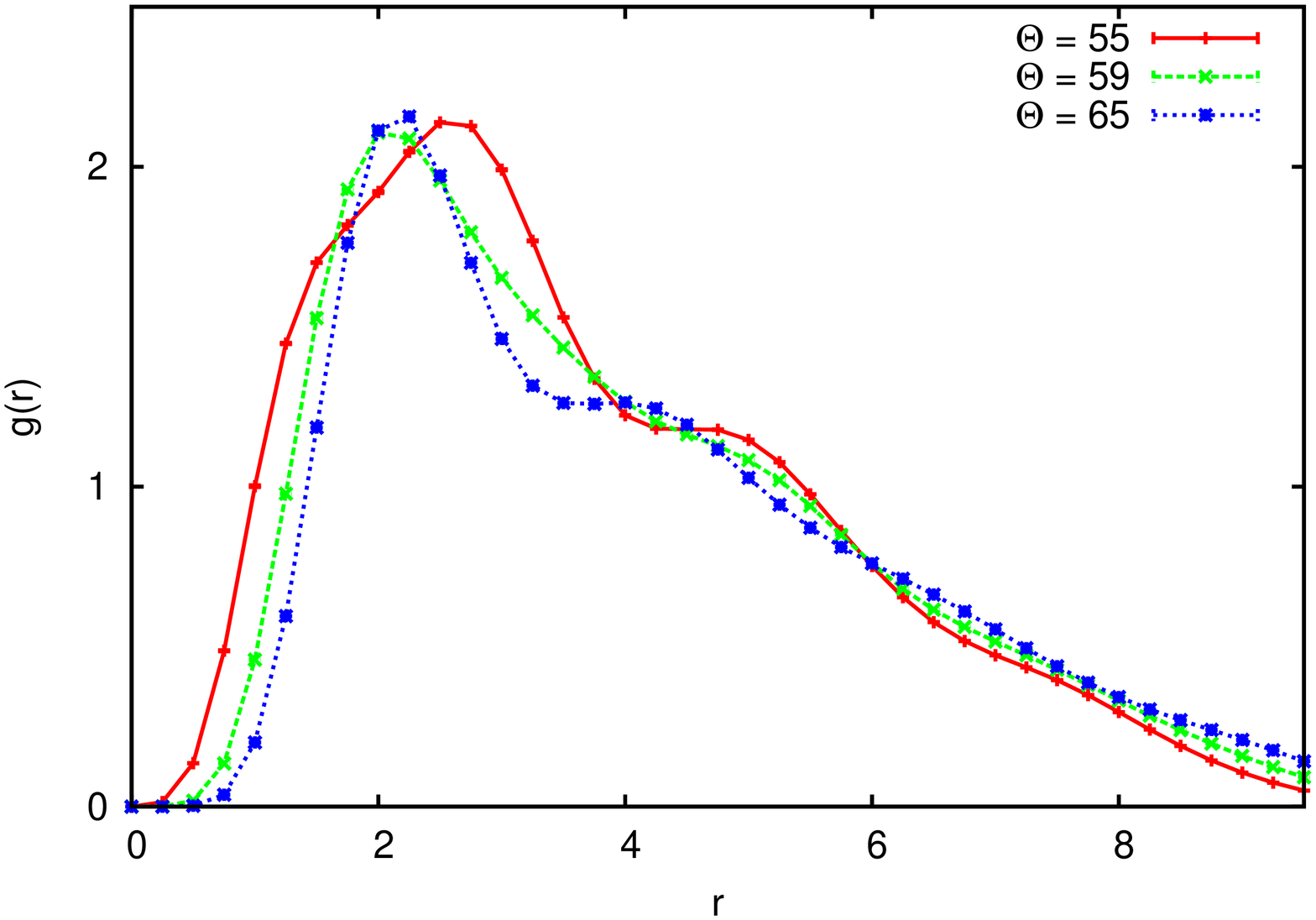}
\includegraphics[angle=270,width=\columnwidth]{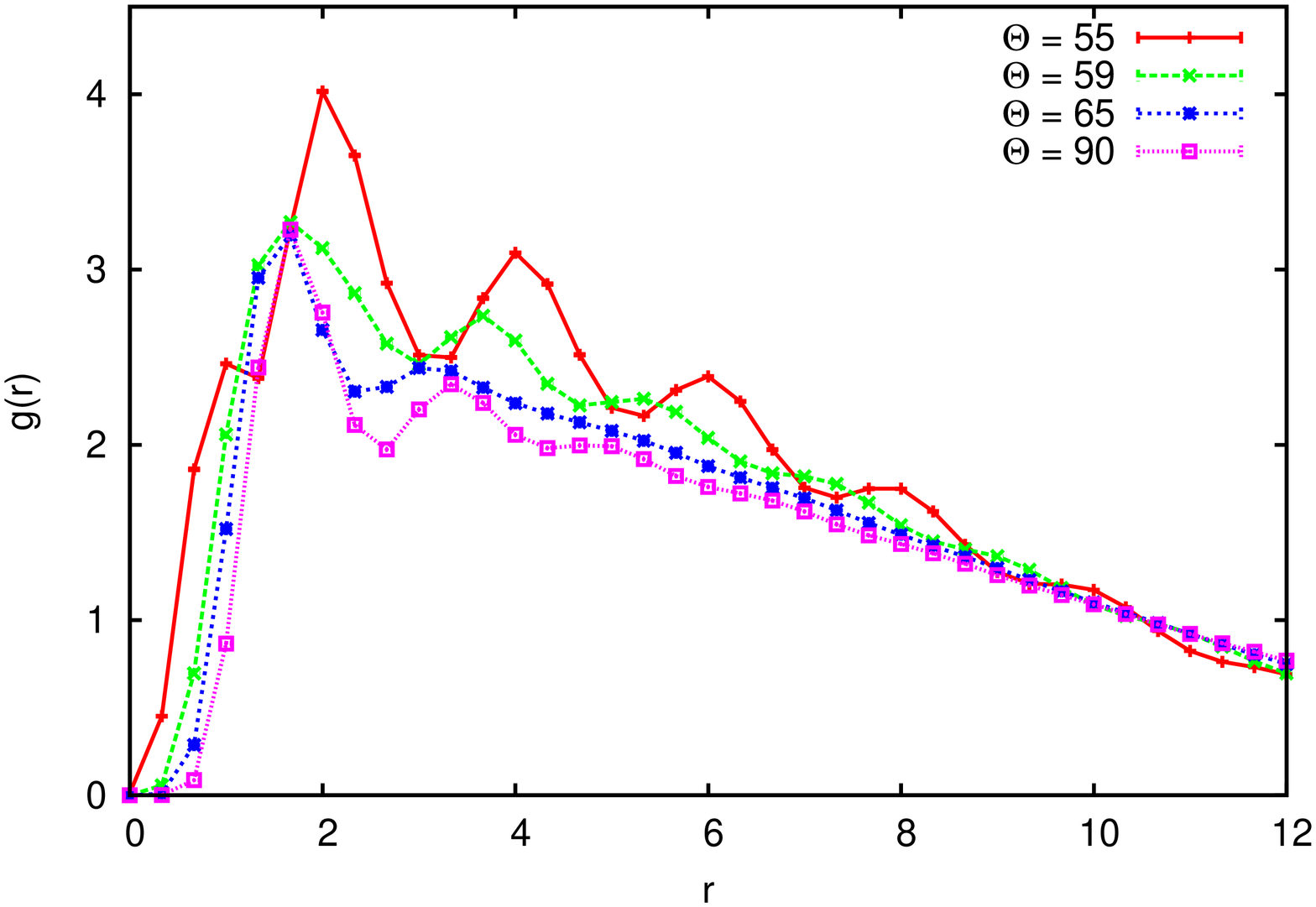}
\caption{\label{gr}Radial correlation function for a $N=19$ and a $N=100$ particle system with increasing $\Theta$.}
\end{figure}

In larger systems we observe a structure displaying shells for vertical dipoles and stripes for smaller $\Theta$. We show in figure (\ref{ro_19}) and (\ref{ro_100}) the density profiles for 19 and 100 atoms.
When we have vertical dipoles the atom are held together toward the centre of the trap. Since this system is symmetric under rotations instead of a lattice in the density plot we see that there are several shells at fixed distances from the centre of the trap. The shells are less evident in the largest system, in which the atoms appear to be more localized.
Tilting the dipoles has the effect of decreasing the repulsion between the dipoles and also of inducing anisotropy. The anisotropy favours the alignment of the atoms along the $x$ direction, and this, along with the localization induced by the trap, causes the dipoles to arrange in stripes. We note that this behaviour is analogue to the one of bulk systems \cite{boronat_bulk_2}. Increasing the angle $\Theta$ has the effect of disrupting the alignment, and the system passes gradually from a structure characterized by stripes to one characterized by shells. By looking at the radial correlation functions, figure (\ref{gr}), we can see that the particle localization is stronger when the dipoles are very tilted and when they are vertical. On the other hand the localization is weaker at intermediate $\Theta$. We will see that this affects the superfluidity of the system.

\begin{figure}
\includegraphics[angle=270,width=\columnwidth]{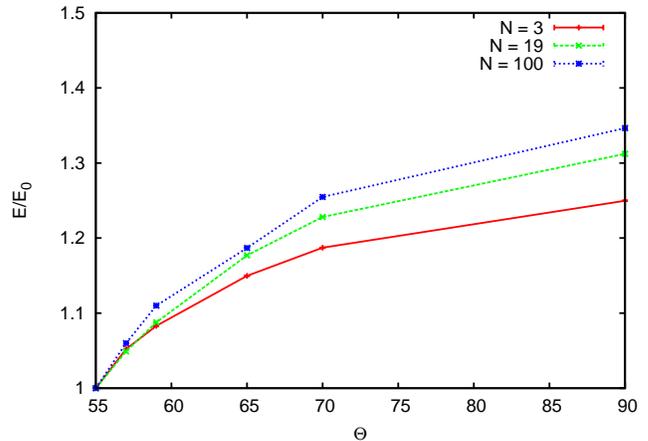}
\caption{\label{energy}Energy per particle as a function of the tilt angle $\Theta$ for a 3, 19 and 100 atoms. The energy is rescaled with respect to $E_0=E(\Theta=55^\circ)$. We have $E_0 = 2.939 \pm 0.001$, $E_0 = 11.278 \pm 0.003$ and $E_0 = 32.371 \pm 0.013 $ in units of $\hbar\omega_0$ for $N=3,19 \text{ and } 100$ atoms respectively.}
\end{figure}

We also measured the energy per particle as a function of the tilt angle. Unsurprisingly we see that the energy steadily increases with $\Theta$ (as the dipolar interaction becomes more and more repulsive) and we note that this increase is qualitatively similar for every size of the system that we studied. We note that the energy per particle here depends on the size of the system, as here we studied a small, confined system, and not a bulk system at the thermodynamic limit. We show the energy in figure (\ref{energy}).

\begin{figure}
\includegraphics[angle=270,width=\columnwidth]{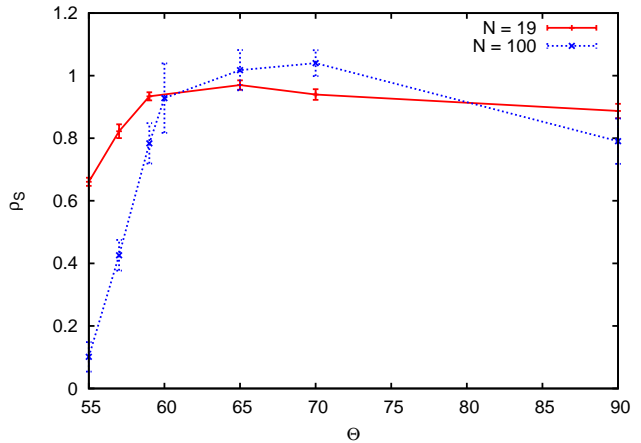}
\caption{\label{super}Superfluid fraction as a function of the tilt angle $\Theta$ for a 19 and 100 atoms system}
\end{figure}

Then we studied the superfluid fraction for the 19 and 100 particle systems. The results are shown in figure (\ref{super}). 
We see that systems of vertical dipoles are superfluid; actually we can argue that the nonclassical inertia along with the shell structure in the particle density can be interpreted as supersolidity; the fact that superfluidity seems to decrease increasing the particle number is in accord with the fact that at the thermodynamic limit the supersolid should not be stable \cite{boninsegni_2}. 

We observe that in strongly tilted systems the superfluidity is suppressed. This effect is especially evident in the larger system. We can also see that the superfluid fraction start from a minimum at $\Theta=55^\circ$, then it rises to a maximum and then it decreases again going towards $\Theta=90^\circ$. We can explain this feature recalling that in a system with vertical dipoles the localization is stronger than in a system with intermediate tilting (see e.g.\ the pair correlation functions $g(r)$ in figure \ref{gr}). This localization has the effect of hindering particle exchanges, thus decreasing the superfluidity of the sample.

We finally compare the quantum system with its classical analogue, to understand the effects of quantum and thermal fluctuations on the trapped dipoles.
We obtained ground state estimates using the simulated annealing technique, and we performed classical Monte Carlo simulations at different temperatures evaluating the single particle density $\rho(x,y)$. We performed these computations for the 19 and 100 particle systems.
\begin{figure}
\includegraphics[angle=270,width=0.49\columnwidth]{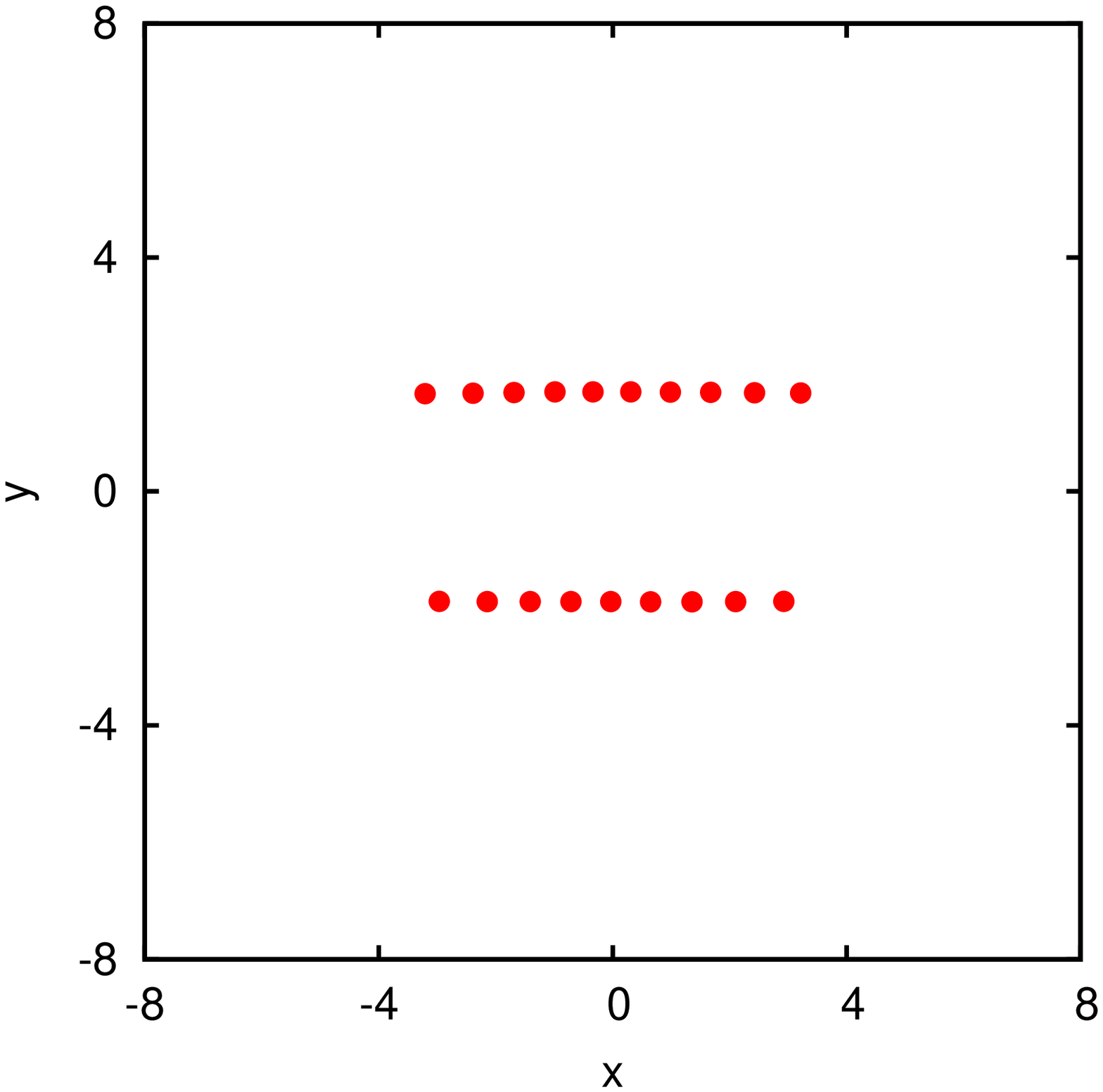}
\includegraphics[angle=270,width=0.49\columnwidth]{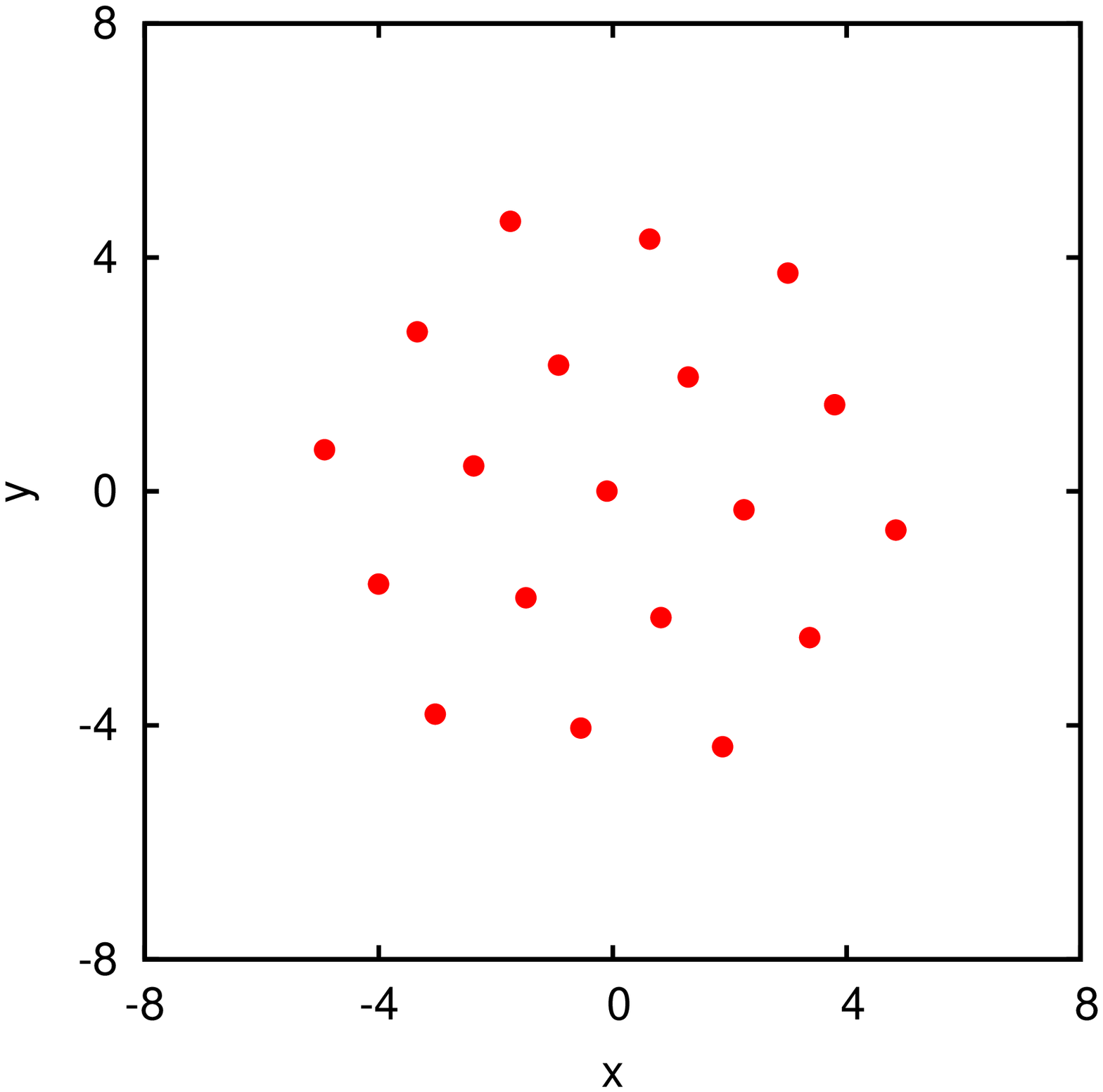}
\caption{\label{gs_19}Classical ground state for a 19 particles system with $\Theta=55^\circ$ (left) and $\Theta=90^\circ$ (right).}
\end{figure}
\begin{figure}
\includegraphics[angle=270,width=0.49\columnwidth]{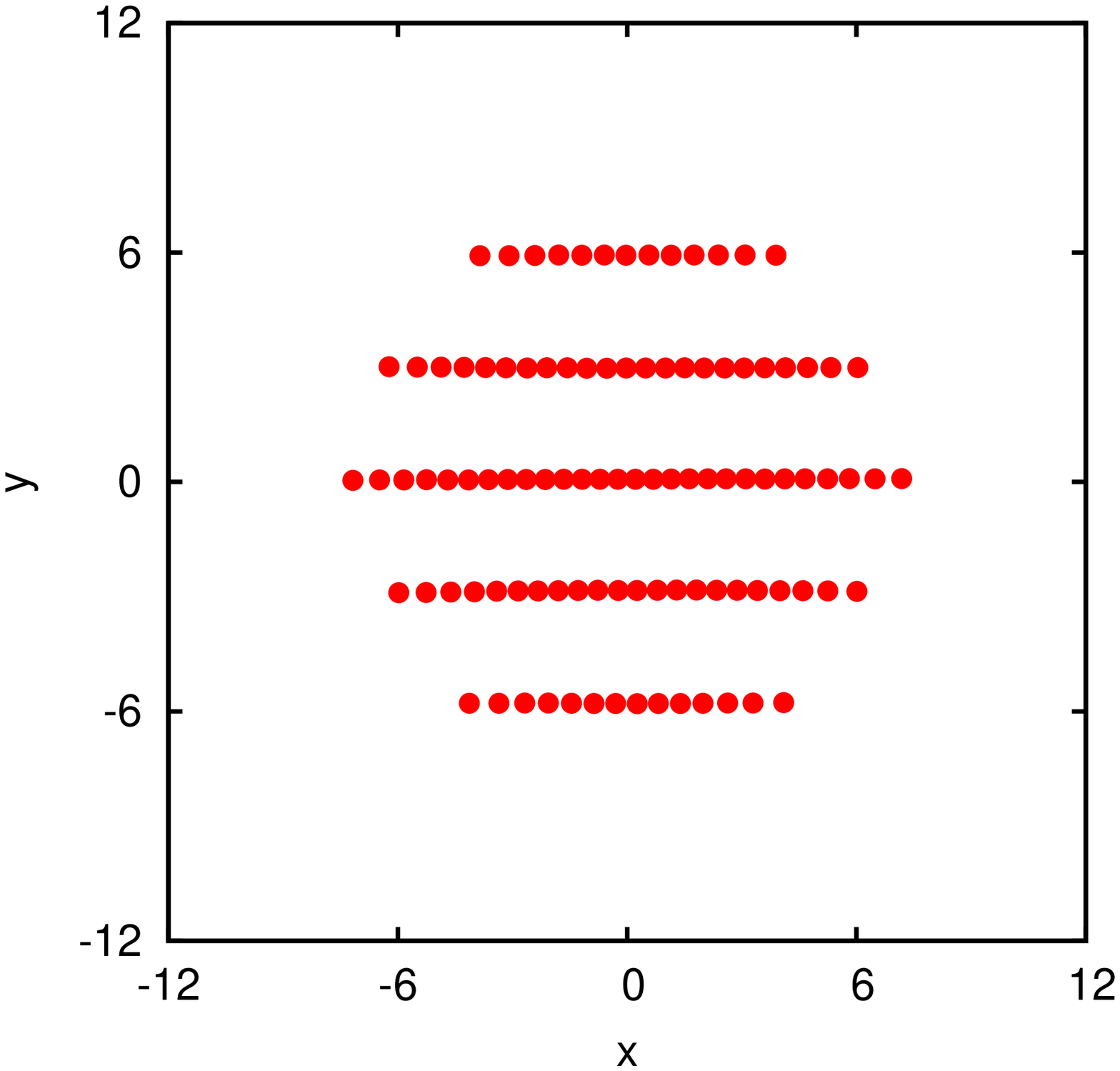}
\includegraphics[angle=270,width=0.49\columnwidth]{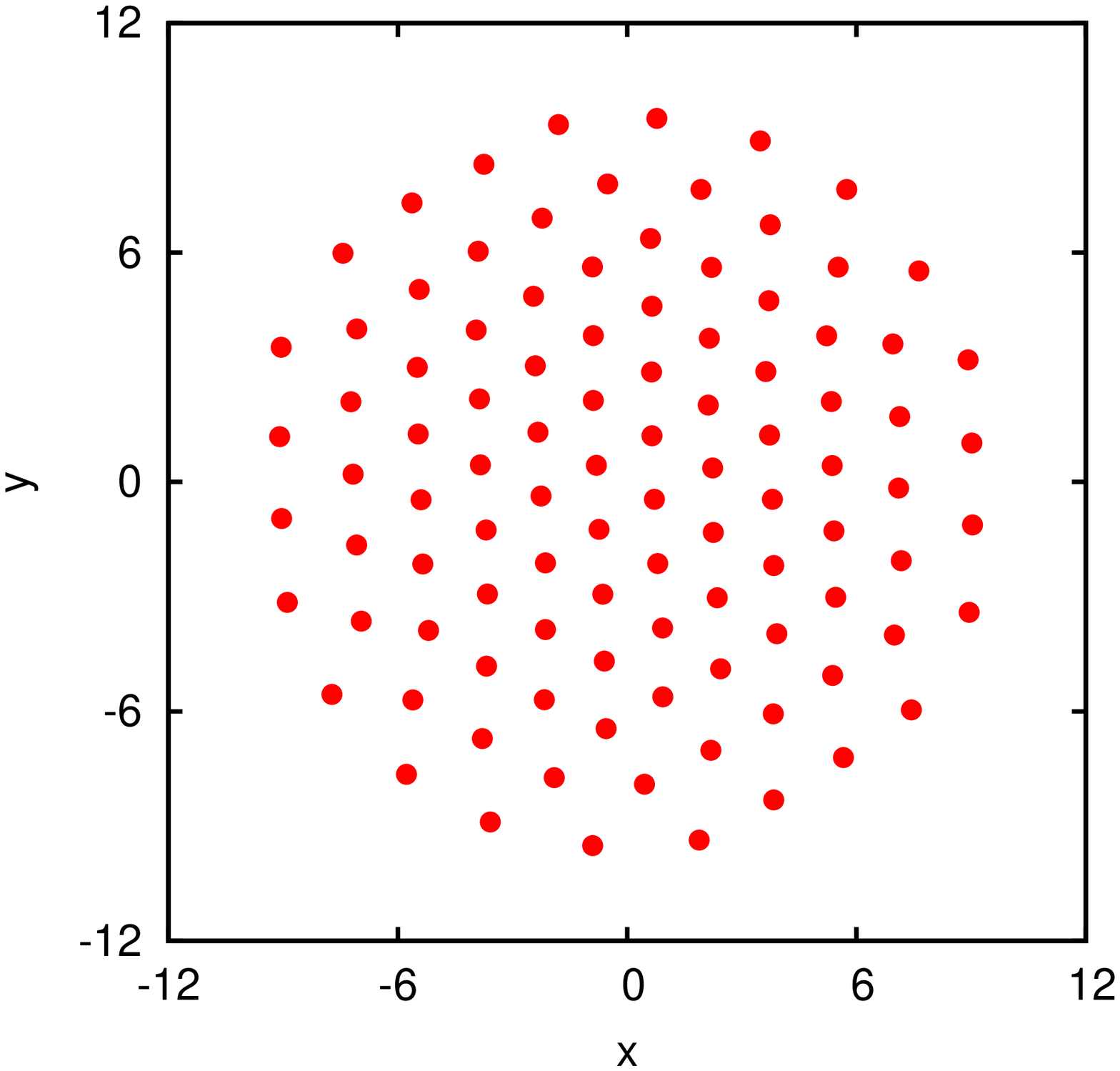}
\caption{\label{gs_100}Classical ground state for a 100 particles system with $\Theta=55^\circ$ (left) and $\Theta=90^\circ$ (right).}
\end{figure}

\begin{figure}
\includegraphics[angle=270,width=0.32\columnwidth]{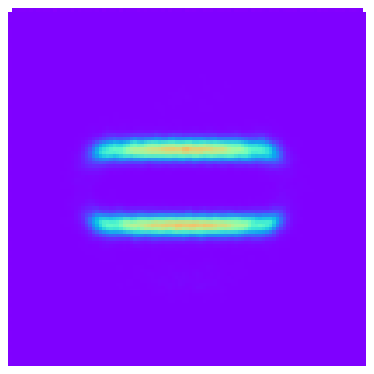}
\includegraphics[angle=270,width=0.32\columnwidth]{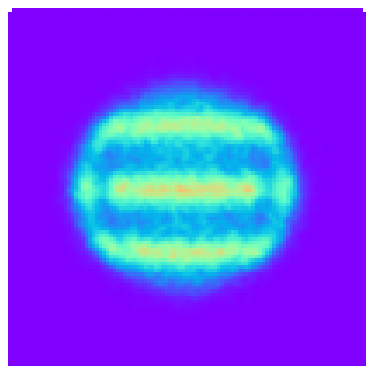}
\includegraphics[angle=270,width=0.32\columnwidth]{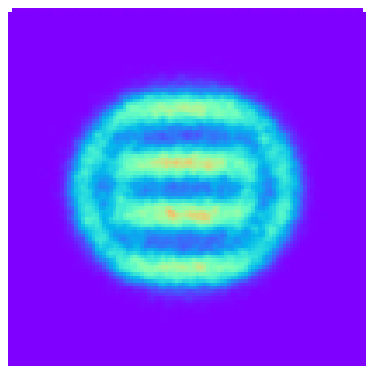}
\includegraphics[angle=270,width=0.32\columnwidth]{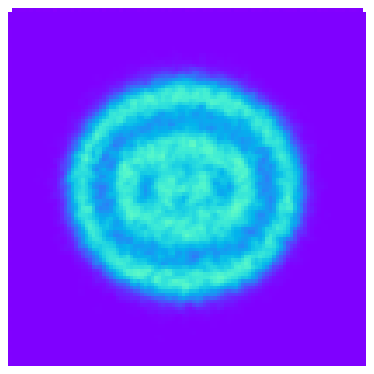}
\includegraphics[angle=270,width=0.32\columnwidth]{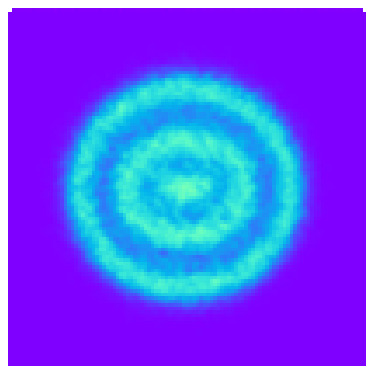}
\includegraphics[angle=270,width=0.32\columnwidth]{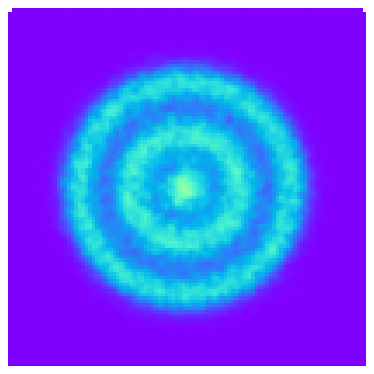}
\caption{\label{classical_19}Particle density for a 19 particles classical system with increasing $\Theta$. We have respectively $\Theta=55^\circ$, $\Theta=57^\circ$, $\Theta=59^\circ$, $\Theta=65^\circ$, $\Theta=70^\circ$, $\Theta=90^\circ$. The temperature is $T=0.6$.}
\end{figure}

\begin{figure}
\includegraphics[angle=270,width=0.32\columnwidth]{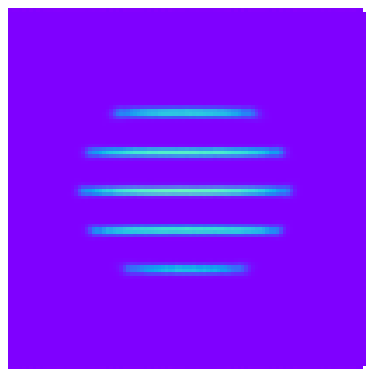}
\includegraphics[angle=270,width=0.32\columnwidth]{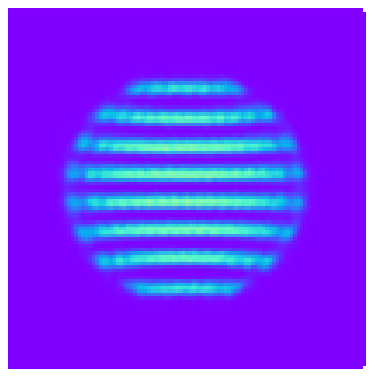}
\includegraphics[angle=270,width=0.32\columnwidth]{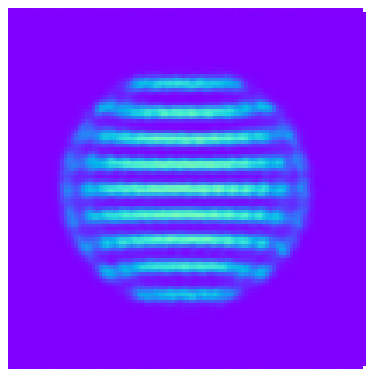}
\includegraphics[angle=270,width=0.32\columnwidth]{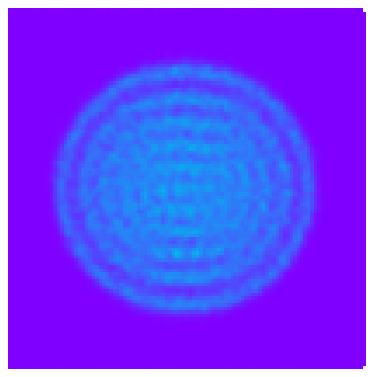}
\includegraphics[angle=270,width=0.32\columnwidth]{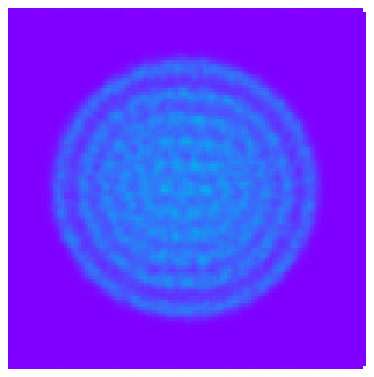}
\includegraphics[angle=270,width=0.32\columnwidth]{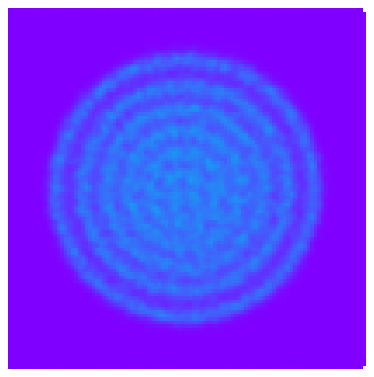}
\caption{\label{classical_100}Particle density for a 100 particles classical system with increasing $\Theta$. We have respectively $\Theta=55^\circ$, $\Theta=57^\circ$, $\Theta=59^\circ$, $\Theta=65^\circ$, $\Theta=70^\circ$, $\Theta=90^\circ$. The temperature is $T=0.6$.}
\end{figure}

\begin{figure}
\includegraphics[angle=270,width=\columnwidth]{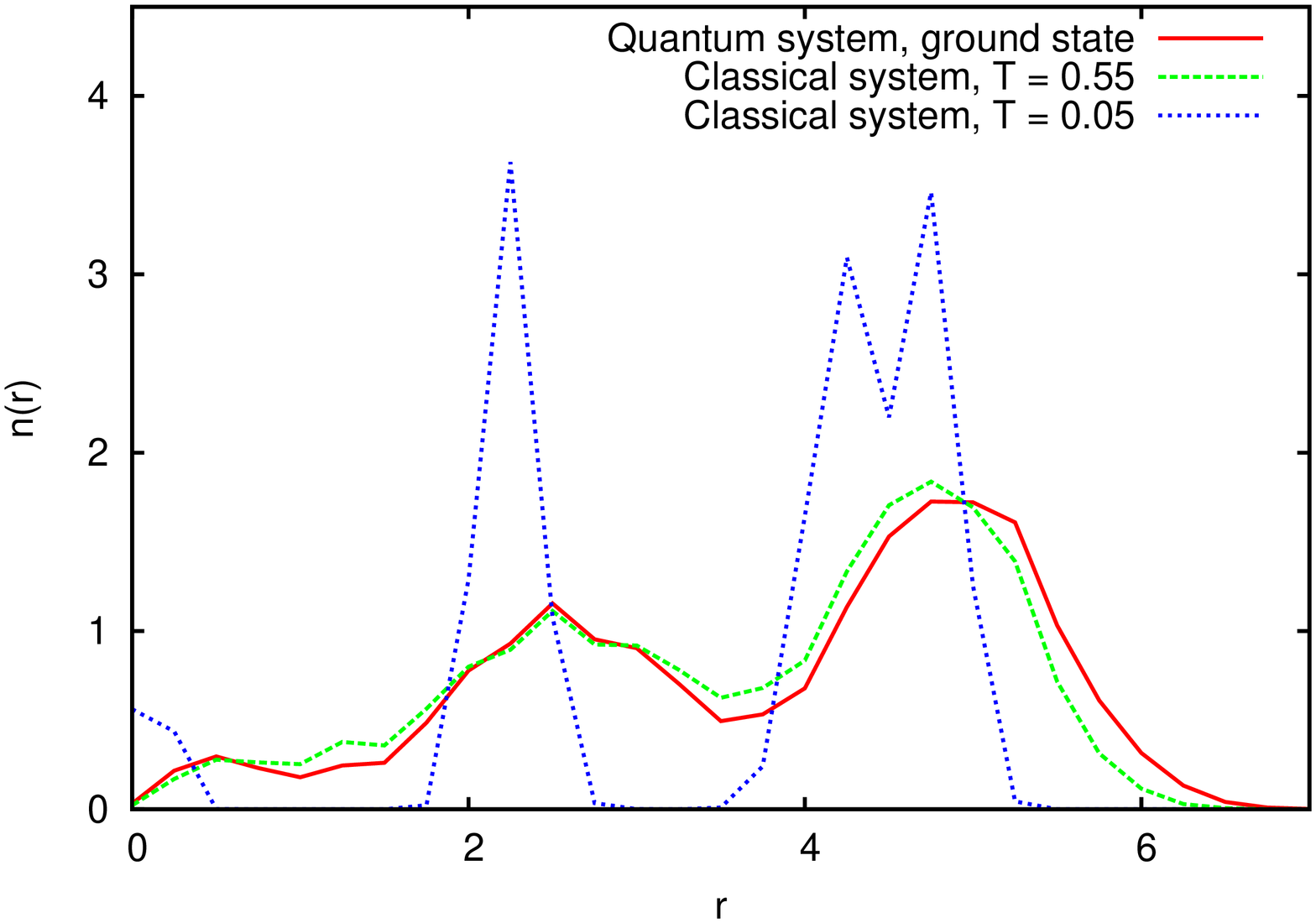}
\includegraphics[angle=270,width=\columnwidth]{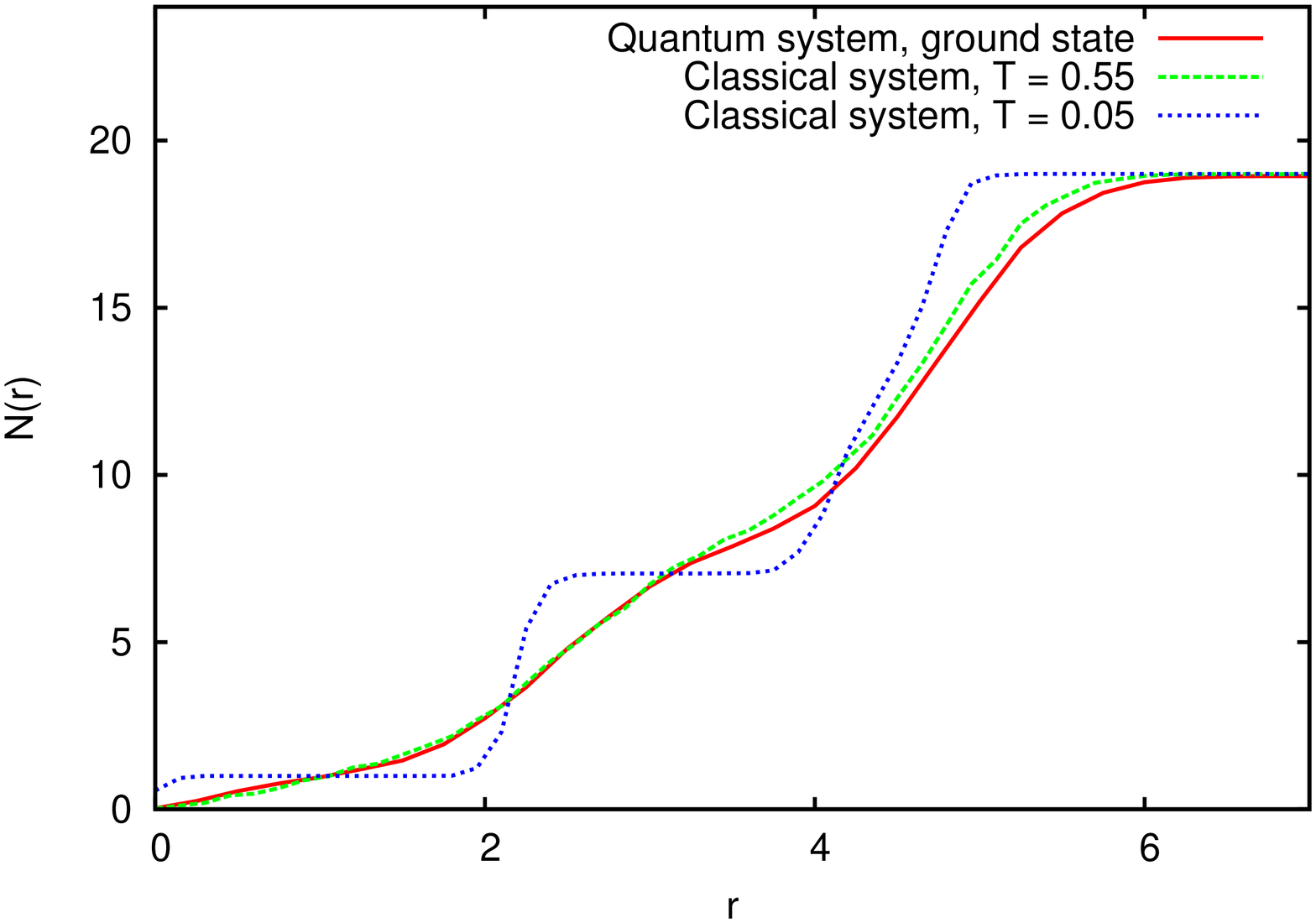}
\caption{\label{quant_class} Radial density (left) and total number of particle contained in a circle of radius $r$ (right) for a quantum system (red) and a classical one at $T=0.55$ (green) and $T=0.05$ (blue). On the right we have lines at $N=1$, $N=7$,$N=8$ and $N=19$ to emphasize the particle numbers corresponding to relevant closed shells.}
\end{figure}

As a general remark we found that a finite temperature for $\Theta$ large enough the classical system behaves quite similarly to the quantum one. 
In the 19 particle quantum system with vertical dipoles ($\Theta=90^\circ$) we see that in the classical ground state we have a triangular lattice, with an atom in the center of the trap, a shell made by 6 atoms and finally a third shell with 12 atoms. In the quantum system the situation is slightly different: in the ground state we can observe a configuration with 1-7-11 atoms per shell. We have an analogue situation for the classical system at finite temperature: as we can see from the radial density in figure \ref{quant_class} the quantum system displays a behaviour which is very similar to the classical one for $T=0.55$, with the difference that here the fluctuations aren't due to thermal fluctuations but to the zero point motion. For very low temperatures instead the ground state lattice is not especially perturbed, and the radial density displays sharper peaks. We note that the difference between the structures displayed by the quantum and classical systems depends on the strength of the interaction between the dipoles as shown in \cite{boninsegni_2}.

\begin{figure}
\includegraphics[angle=270,width=0.32\columnwidth]{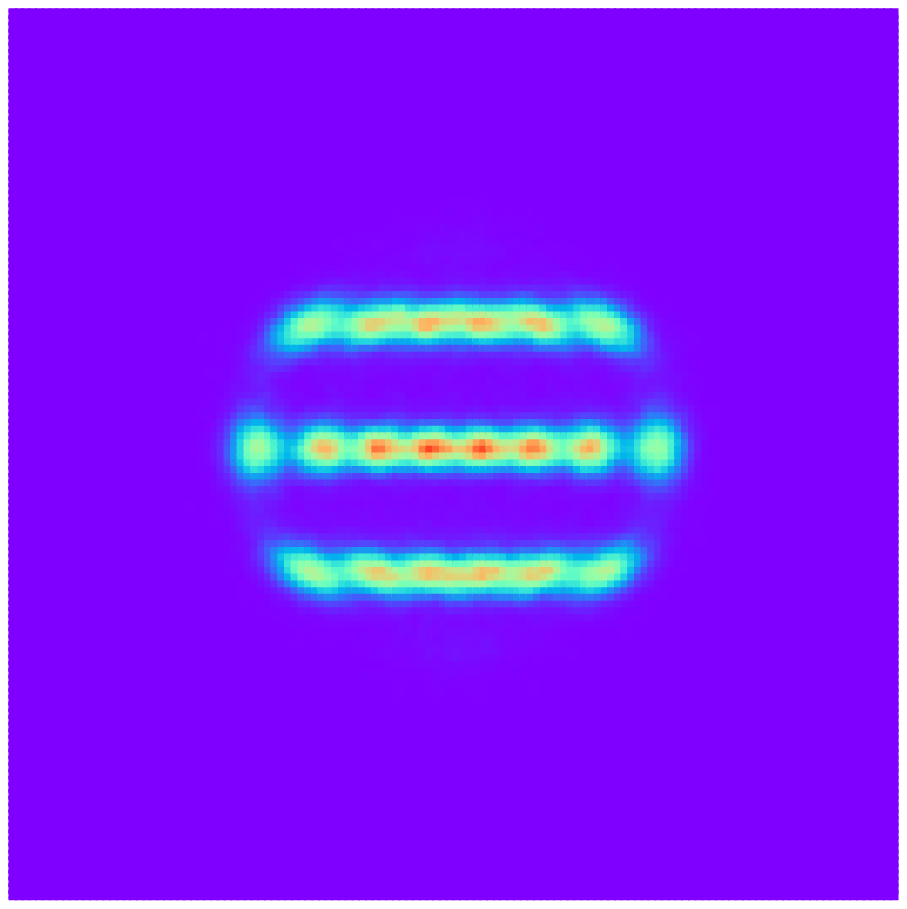}
\includegraphics[angle=270,width=0.32\columnwidth]{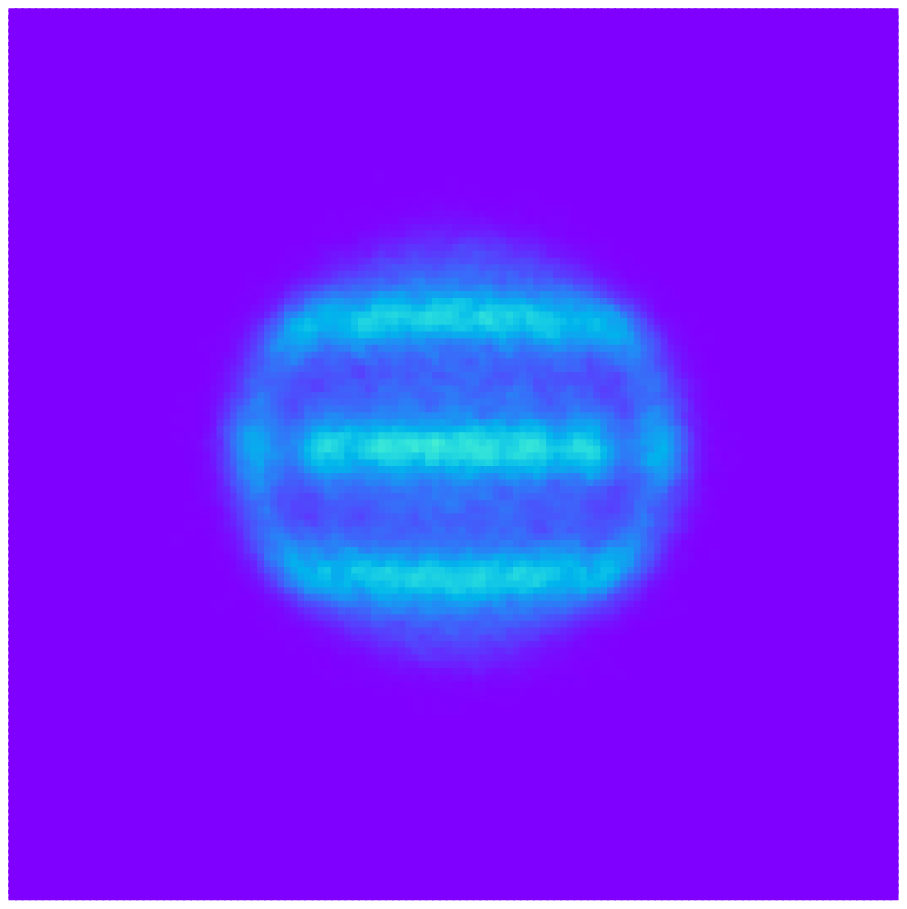}
\includegraphics[angle=270,width=0.32\columnwidth]{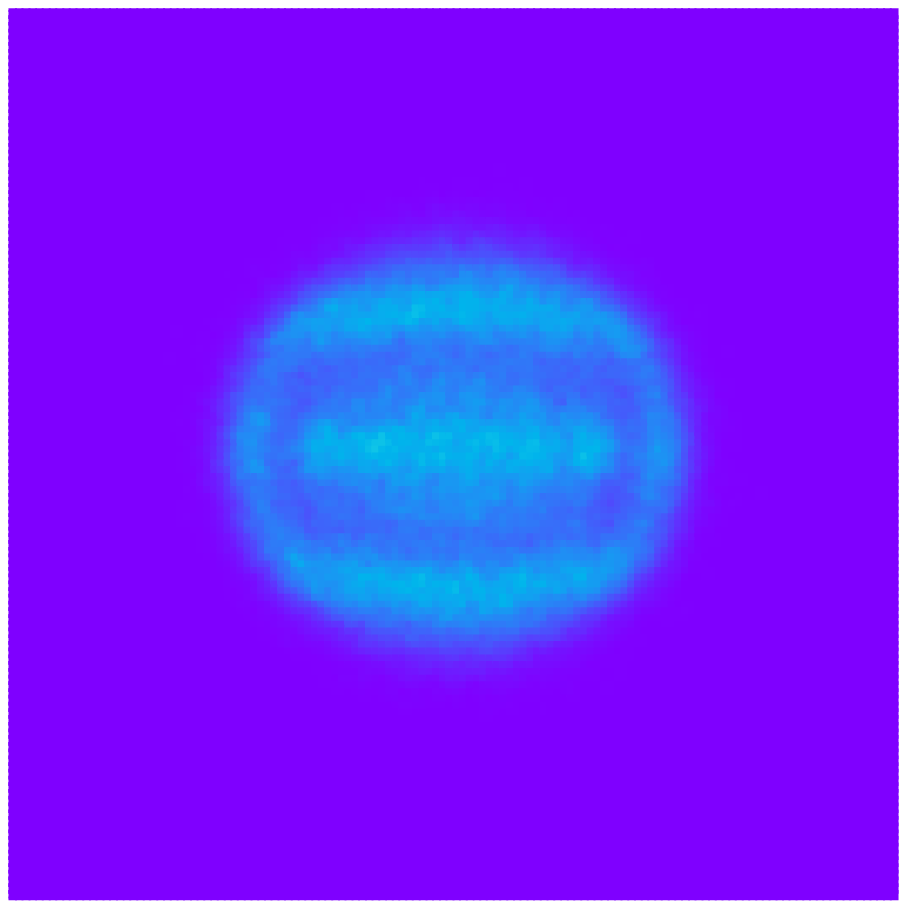}
\includegraphics[angle=270,width=0.32\columnwidth]{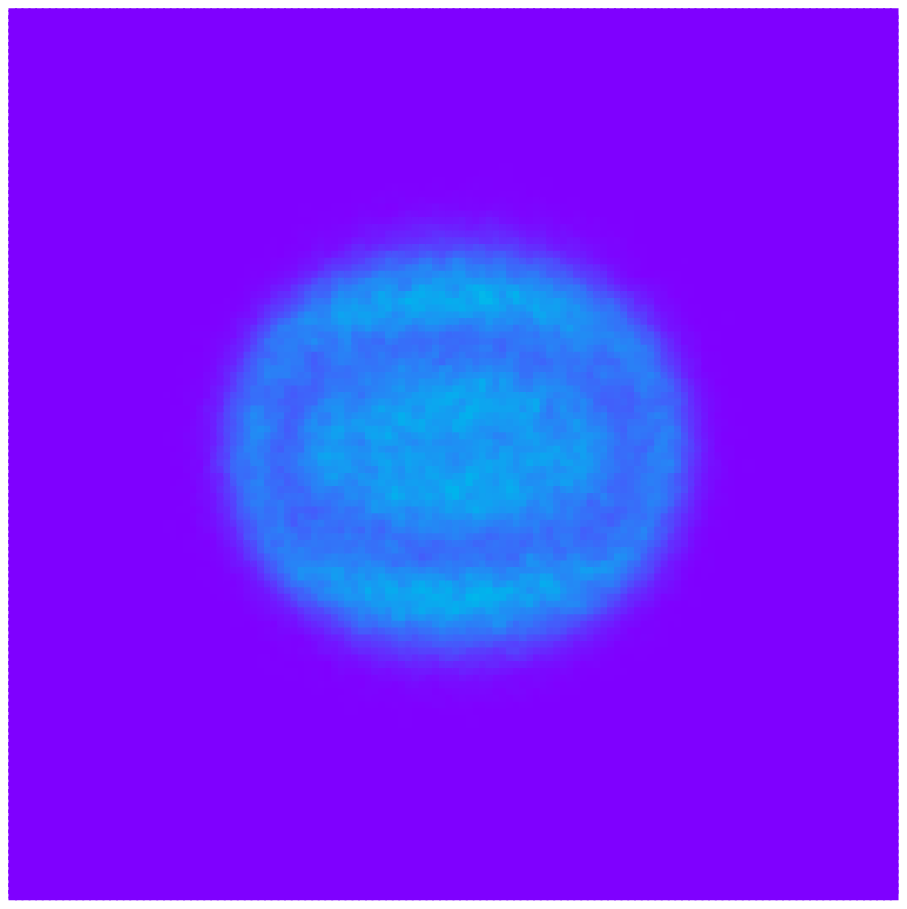}
\includegraphics[angle=270,width=0.32\columnwidth]{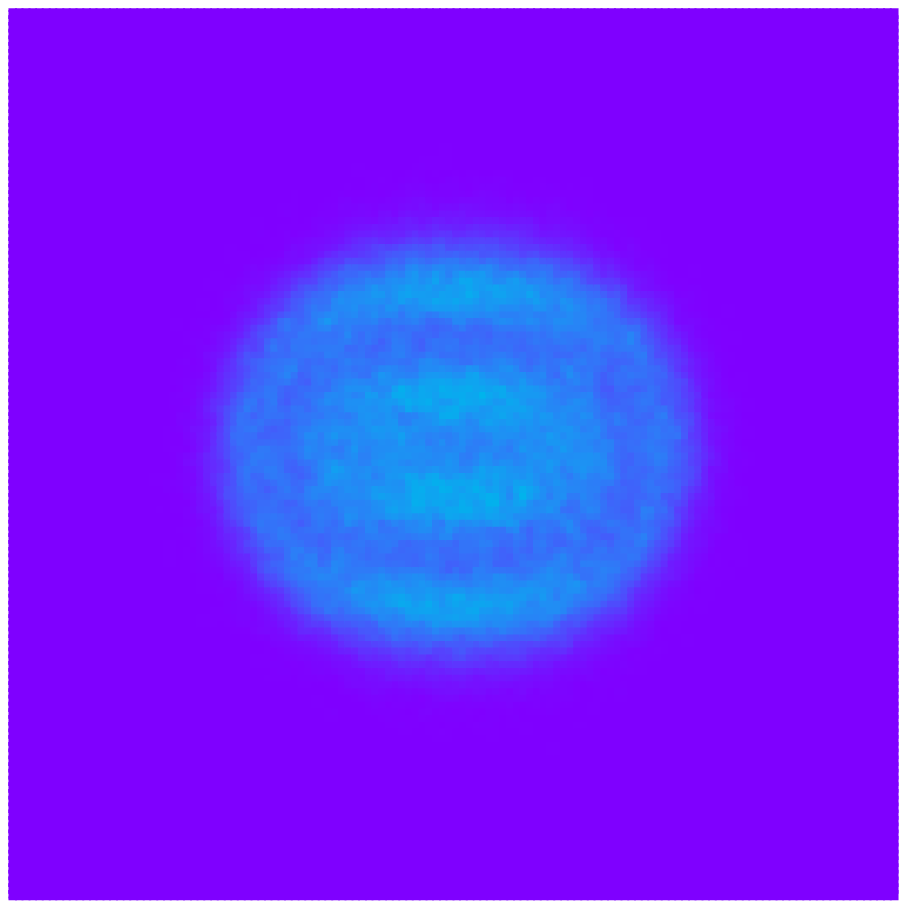}
\includegraphics[angle=270,width=0.32\columnwidth]{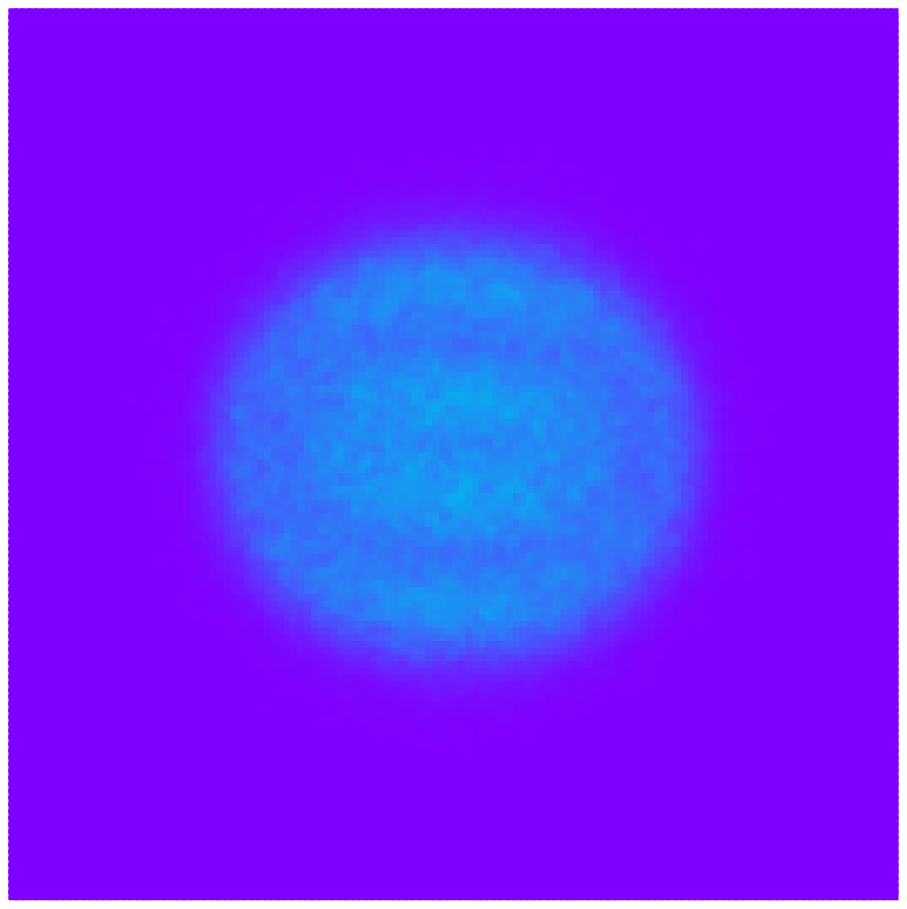}
\caption{\label{wasd}Particle density for a 19 particles classical system with increasing temperature with $\Theta=57^\circ$. We show results for $T=0.4$, $T=0.6$, $T=0.7$, $T=0.8$, $T=1.0$ and $T=1.2$.}
\end{figure}

The temperature at which the classical system behaves like the quantum one seems to increase as $\Theta$ gets smaller. We show an example for $\Theta=57^\circ$ in figure \ref{wasd}; here at low temperatures instead of seeing 4 stripes as in the quantum system we notice that the system displays just 3; increasing the temperature has the eventual effect of splitting the central stripe, recovering a configuration that is more similar to the quantum case. At higher temperatures the stripes are starting to fade, so that is difficult to recover a full similarity with the quantum system.

Analogue behaviour can also be found in the larger system. We show in figure (\ref{gs_100} and (\ref{classical_100}) the results. We note again that vertical dipoles in the ground state form a triangular lattice. There are some defects, likely due to the trapping potential and to the fact that 100 is not a number compatible with closed shells for a triangular lattice; these defects are pushed on the border of the crystal, while the inner structure is essentially crystalline. 
Observing the particle density, we can see again the systems are quite different at low $\Theta$: the stripes in the classical system are straight, while in the quantum system they are somewhat bent. This difference again becomes less evident increasing $\Theta$ and $T$.

\section{Conclusions}
From our simulations we could see that the arrangement of the atoms in the optical trap changes tilting the dipole moments. Vertical dipoles tend to form shells around the center of the trap, while tilting the dipoles favours a striped structure. The stripes can be observed if the tilting is quite large: the stripes appear in fact with $\Theta \lesssim 60$, while with $\Theta \gtrsim 65$ the shell structure is essentially retrieved.

This behaviour of the quantum system is close to the behaviour of its classical analogue at a finite temperature; we can argue that the structures arising in the trap are due to the form of the potential and the presence of fluctuations, without especially meaningful distinctions between zero point or thermal ones; we note however that for very tilted dipoles the temperature for which the behaviour of the two systems is similar becomes higher.

This system displays superfluidity; superfluidity is suppressed by a strong tilting in the dipoles, and it is actually maximum at intermediate tilting, when the atomic localization is weaker. The shell structure along with the superfluidity in the system with vertical dipoles can be interpreted as a signature of supersolidity.

\section*{Acknowledgements}
The author thanks Saverio Moroni for the useful discussions.

\end{document}